\documentclass[aps,prb,twocolumn,longbibliography,superscriptaddress]{revtex4-1}

\usepackage{graphicx}% Include figure files
\usepackage{dcolumn}% Align table columns on decimal point
\usepackage{bm}% bold math
\usepackage[version=3]{mhchem}
\usepackage{color}

\begin{document}

\title{Electronic structure and properties of lithium-rich complex oxides}
\author{Khang Hoang}
\affiliation{Center for Computationally Assisted Science and Technology, North Dakota State University, Fargo, North Dakota 58108, United States}
\affiliation{Department of Physics, North Dakota State University, Fargo, North Dakota 58108, United States}
\author{Myungkeun Oh}
\affiliation{Materials and Nanotechnology Program, North Dakota State University, Fargo, North Dakota 58105, United States}
\author{Yongki Choi}
\email{yongki.choi@ndsu.edu}
\affiliation{Department of Physics, North Dakota State University, Fargo, North Dakota 58108, United States}
\affiliation{Materials and Nanotechnology Program, North Dakota State University, Fargo, North Dakota 58105, United States}

\date{\today}

\begin{abstract}

Lithium-rich complex transition-metal oxides Li$_2$MoO$_3$, Li$_2$RuO$_3$, Li$_3$RuO$_4$, Li$_3$NbO$_4$, Li$_5$FeO$_4$, Li$_5$MnO$_4$ and their derivatives are of interest for high-capacity battery electrodes. Here, we report a first-principles density-functional theory study of the atomic and electronic structure of these materials using the Heyd-Scuseria-Ernzerhof (HSE) screened hybrid functional which treats all orbitals in the materials on equal footing. Dimerization of the transition-metal ions is found to occur in layered Li$_2$MoO$_3$, in both fully lithiated and partially delithiated compounds. The Ru--Ru dimerization does not occur in fully lithiated Li$_2$RuO$_3$, in contrast to what is commonly believed; Ru--Ru dimers are, however, found to occur in the presence of lithium vacancies caused by lithium loss during synthesis and/or lithium removal during use. We also analyze the electronic structure of the complex oxides and discuss the delithiation mechanism in these battery electrode materials.

\end{abstract}

\pacs{}
%\keywords{}

\maketitle


\section{Introduction}\label{sec;intro}

Lithium-rich complex transition-metal oxides have been of great interest for lithium-ion battery electrodes due to their high theoretical capacity. These materials include layered oxides Li$_2$MoO$_3$,\cite{James1988JSSCMo,Ma2014CM} Li$_2$RuO$_3$,\cite{James1988JSSC,Johannes2008PRB} and Li$_3$RuO$_4$,\cite{Jacquet2017CM} and anti-fluorite Li$_5$FeO$_4$\cite{Zhan2017NE} as well as their derivatives.\cite{Self2018CM,Sathiya2013,Sathiya2013NM,Sathiya2015NM,Assat2017CM} It has been reported that some of these battery materials exhibit both cationic and anionic redox behavior.\cite{Assat2018NE,Zhan2017NE} An understanding of the delithiation mechanism would necessarily require a detailed knowledge of the materials' electronic structure.\cite{Hoang2018JPCM} On another fundamental aspect, metal--metal bond disproportionation has been reported to occur on the transition-metal sublattice in Li$_2$MoO$_3$ and Li$_2$RuO$_3$,\cite{James1988JSSCMo,Takahashi2008JPCS,Ma2014CM} the two layered oxides with a honeycomb transition-metal network. The phenomenon is, however, not well understood. Previous reports on the phenomenon have been conflicting and indicated that the occurrence of the Ru--Ru dimerization may be dependent on the synthesis procedure.\cite{Wang2014PRB,Jimenez-Segura2016PRB}

Computational studies of the above mentioned materials have been carried out by different research groups,\cite{Baldoni2013JMCA,Tian2017PCCP,Johannes2008PRB,Pchelkina2015PRB,Saubanere2016EES,Wang2014PRB,Jacquet2017CM} often using first-principles calculations based on density-functional theory (DFT) within the local-density (LDA) or generalized gradient (GGA) approximation\cite{LDA1980,PW91} and/or the DFT+$U$ extension\cite{anisimov1991} where $U$ is the on-site Coulomb correction. There are, however, limitations with these methods when applied to complex transition-metal oxides. It is well known that LDA and GGA tend to overdelocalize electrons and often fail in localized electron systems. The DFT+$U$ method, on the other hand, requires {\it a priori} knowledge of the $U$ parameter. Furthermore, DFT+$U$ calculations with $U$ applied only on the transition-metal $d$ orbitals leave the oxygen $p$ states uncorrected; as a result, the calculations may not be able to reproduce the correct physics, especially in materials where there is strong mixing between the transition-metal $d$ and oxygen $p$ states and/or when the oxygen $p$ states can play an important role.\cite{Hoang2015PRA}  

We herein report a first-principles study of the Li-rich complex oxides using a hybrid DFT/Hartree-Fock method\cite{Perdew1996JCP} in which all electronic states in the materials are treated on equal footing. The focus of this work is on the electronic structure, particularly the nature of the electronic states near the band edges, in the different materials, and its implications on the delithiation mechanism. The dimerization of the transition-metal ions in layered oxides Li$_2$MoO$_3$ and Li$_2$RuO$_3$ is also discussed. 

\section{Methods}\label{sec;method} 

Our calculations are based on DFT, using the hybrid functional of Heyd, Scuseria, and Ernzerhof (HSE),\cite{heyd:8207} the projector augmented wave (PAW) method,\cite{PAW1} and a plane-wave basis set, as implemented in the Vienna {\it Ab Initio} Simulation Package (\textsc{vasp}).\cite{VASP2} The Hartree-Fock mixing parameter ($\alpha$) and the screening length are set to their standard values, 0.25 and 10 {\AA}, respectively, unless otherwise noted. The plane-wave basis-set cutoff is set to 500 eV and spin polarization is included. 

\begin{figure*}%[t]%
\vspace{0.2cm}
\includegraphics*[width=0.85\linewidth]{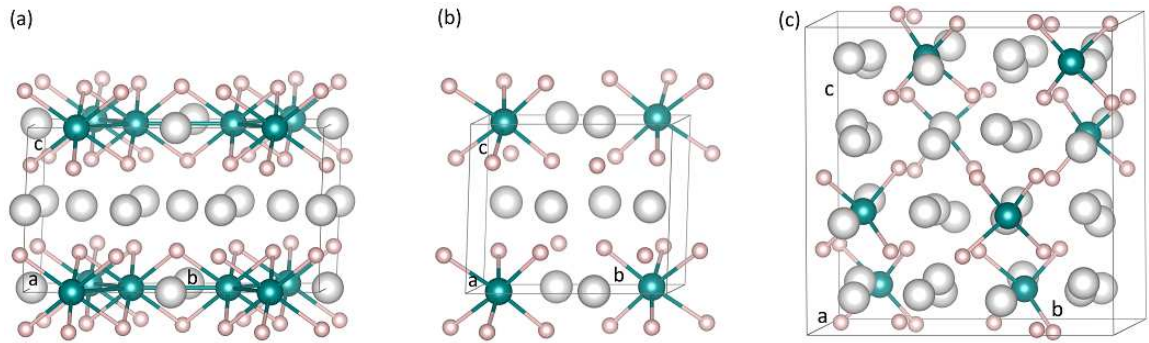}
\caption{Atomic structure of select Li-rich complex oxides: (a) Li$_2$RuO$_3$ (monoclinic, $C2/m$), (b) Li$_3$RuO$_4$ (monoclinic, $P2/a$), and (c) Li$_5$FeO$_4$ (orthorhombic, $Pbca$). Large (gray) spheres are Li, medium (blue) are Ru or Fe, and small (red) are O.}
\label{fig;struct}
\end{figure*}

\begin{table*}%[ht]
\small
\caption{\ Lattice parameters and band gaps ($E_g$) of Li-rich complex oxides, obtained in HSE calculations}\label{lab;latt}
\begin{tabular*}{\textwidth}{@{\extracolsep{\fill}}llllll}
\hline
&&\multicolumn{2}{c}{Calculated}&\multicolumn{1}{c}{Experimental}& \multicolumn{1}{c}{$E_g$} \\
\hline
Li$_2$MoO$_3$& $P1$& AF &$a=4.972$ {\AA}, $b=4.960$ {\AA}, $c=5.224$ {\AA},&& 1.58 eV\\
						 &&&$\alpha=98.24^{\circ}$, $\beta=108.58^{\circ}$, $\gamma=62.80^{\circ}$& \\
						 &&FM & $a=4.998$ {\AA}, $b=4.994$ {\AA}, $c=5.227$ {\AA},&& 1.12 eV\\
						 &&&$\alpha=98.18^{\circ}$, $\beta=108.63^{\circ}$, $\gamma=62.33^{\circ}$& \\
Li$_2$RuO$_3$& $C2/m$& FM & $a=5.105$ {\AA}, $b=8.901$ {\AA}, $c=5.095$ {\AA},&$a=5.021$ {\AA}, $b=8.755$ {\AA}, $c=5.119$ {\AA},& 1.27 eV\\
		 &&&$\beta=109.12^{\circ}$&$\beta=108.95^{\circ}$ (Ref.~\citenum{Wang2014PRB})& \\
Li$_3$RuO$_4$& $P2/a$& FM & $a=5.077$ {\AA}, $b=5.857$ {\AA}, $c=5.121$ {\AA},& $a=5.085$ {\AA}, $b=5.872$ {\AA}, $c=5.125$ {\AA},& 2.11 eV\\
		 &&&$\beta=110.29^{\circ}$&$\beta=110.21^{\circ}$ (Ref.~\citenum{Jacquet2017CM})\\
Li$_3$NbO$_4$& $I\bar{4}3m$ & FM & $a=b=c=8.435$ {\AA} & $a=b=c=8.442$ {\AA} (Ref.~\citenum{Jacquet2017CM})& 5.39 eV\\
Li$_5$FeO$_4$& $Pbca$&FM & $a=9.173$ {\AA}, $b=9.153$ {\AA}, $c=9.114$ {\AA}& $a=9.218$ {\AA}, $b=9.213$ {\AA}, $c=9.159$ {\AA} (Ref.~\citenum{Luge1984}) & 4.41 eV\\
Li$_5$MnO$_4$& $Pbca$ & FM & $a=8.686$ {\AA}, $b=9.348$ {\AA}, $c=9.322$ {\AA}&& 2.25 eV\\
    \hline
  \end{tabular*}
\end{table*}

Calculations for bulk Li$_2$MO$_3$ (two formula units per unit cell), Li$_3$MO$_4$ (two formula units per unit cell), or Li$_5$FeO$_4$ (eight formula units per unit cell) are carried out using a $\Gamma$-centered 8$\times$8$\times$7, 7$\times$7$\times$6, or 4$\times$4$\times$4 {\it k}-point mesh. The experimental atomic structures of Li$_3$RuO$_4$, Li$_3$NbO$_4$,\cite{Jacquet2017CM} and Li$_5$FeO$_4$\cite{Luge1984} are used as the initial structures in the calculations; for the other compounds, the initial atomic structures are taken from those in the Materials Project\cite{Jain2013} database. Mixed-metal compounds are created through partial substitution of transition metals in the host compounds. Larger, up to 4$\times$4$\times$1 (192-atom), supercells are also used in the study of bond disproportionation on the transition-metal sublattice and delithiation mechanism. In all calculations, structural optimizations are performed with the HSE functional and the force threshold is chosen to be 0.01 eV/{\AA}.

DFT$+U$ calculations\cite{dudarev1998} based on the the GGA version of Perdew, Burke, and Ernzerhof (PBE),\cite{GGA} hereafter referred to as PBE$+U$, are also carried out for comparison. The effective $U$ value, i.e., $U-J$, varies from 0 to 2 eV.  

\section{Results and Discussion}\label{sec;results}

\subsection{Transition metals on the honeycomb network: To dimerize or not to dimerize?}\label{sec;dimerize}

Figure \ref{fig;struct} shows the atomic structure of select Li-rich complex oxides. The lattice parameters of all the single phases considered in this work are summarized in Table \ref{lab;latt}. The structure of Li$_2$MO$_3$ (M = Mo, Ru) and Li$_3$RuO$_4$ has alternate Li and M/Li layers. In Li$_2$MO$_3$, the transition metal forms a honeycomb network in the M/Li layer [Fig.~\ref{fig;struct}(a)]; in Li$_3$RuO$_4$, Ru forms zigzag chains. In these oxides, as well as in Li$_3$NbO$_4$, the transition metal is octahedrally coordinated with oxygen. The transition metal in Li$_5$MO$_4$ (M = Fe, Mn) is, on the other hand, tetrahedrally coordinated [Fig.~\ref{fig;struct}(c)]. We find that in these complex oxides, except Li$_2$MoO$_3$ (see more below), the antiferromagnetic (AF) and ferromagnetic (FM) spin configurations are almost degenerate in energy.

In layered Li$_2$MoO$_3$, bond disproportionation occurs on the transition-metal network with all Mo ions dimerized. An AF configuration of Li$_2$MoO$_3$ with alternate up and down spins is lower in energy than the FM one by 74 meV per formula unit (f.u.). In this AF configuration, the Mo--Mo dimers have a calculated bond length of 2.45 {\AA}, significantly shorter than the lengths (3.09 {\AA} and 3.17 {\AA}) of the other Mo--Mo bonds. The Mo ions has a calculated local magnetic moment of 0.79$\mu_{\rm B}$, significantly smaller than that expected of Mo$^{4+}$ with two unpaired $d$ electrons. They can thus be regarded as {\it effectively} Mo$^{5+}$. In the FM configuration, the bond length within the Mo--Mo dimer is 2.47 {\AA}, compared to that of 3.15 {\AA} of the other Mo--Mo bonds; half of the Mo ions has a local magnetic moment of 0.45 $\mu_{\rm B}$ and the other half has that of 1.17$\mu_{\rm B}$. Note that the undimerized FM configuration of Li$_2$MoO$_3$ is higher in energy than the dimerized AF one by 262 meV/f.u.; in the former, the Mo--Mo bond lengths are 2.86 {\AA} and 3.11 {\AA}, and the Mo ions has a magnetic moment of 1.68$\mu_{\rm B}$. Clearly, the Mo--Mo dimerization leads to reduced local magnetic moments. This is because some Mo $4d$ electrons participate in the formation of Mo--Mo covalent bonds, which leads to a reduction in the number of electrons contributing to effective localized moments, an effect believed to also occur in other materials with transition-metal dimers.\cite{Streltsov2016PNAS} The bond lengths obtained in our calculations for the low-energy and dimerized configuration of Li$_2$MoO$_3$ are in excellent agreement with the short and long Mo--Mo bonds of 2.524 {\AA} and 3.255 {\AA} observed in experiments.\cite{James1988JSSCMo}

\begin{figure}%[t]%
\vspace{0.2cm}
\includegraphics*[width=0.98\linewidth]{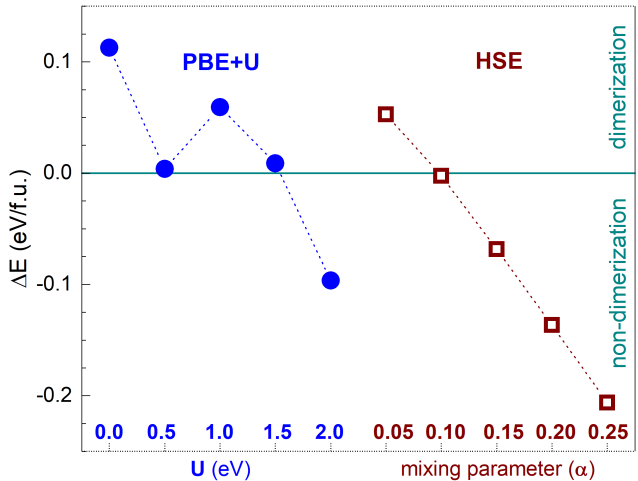}
\caption{Total-energy difference ($\Delta E$) between the dimerization and non-dimerization configurations of fully lithiated Li$_2$RuO$_3$, obtained in PBE$+U$ and HSE calculations with different $U$ and $\alpha$ values. Negative values mean the non-dimerization configuration is lower in energy than the dimerization one. The dotted lines are just to guide the eyes.}
\label{fig;DeltaE}
\end{figure}

Interestingly, Ru--Ru dimerization is not observed in fully lithiated Li$_2$RuO$_3$, except in PBE$+U$ calculations with $U \le 1.5$ eV or in HSE calculations with very small mixing parameter values (e.g., $\alpha < 0.10$); see Fig.~\ref{fig;DeltaE}. The results reported in Fig.~\ref{fig;DeltaE} are obtained in calculations using the large, 4$\times$4$\times$1 (192-atom), supercells to release possible constraints on the crystal symmetry. Note that in the PBE$+U$ calculations, the $U$ term is applied only on the Ru $4d$ orbitals and the O $2p$ states remain uncorrected. In materials such as Li$_2$RuO$_3$ where the valence-band top has a significant contribution from the O $2p$ states and there is strong mixing between the transition-metal $d$ and oxygen $p$ states (see Sec.~\ref{sec;electronic}), the PBE$+U$ calculations would have limited predictive power. Our results show that, for a reasonable choice of the computational method (here, HSE with $\alpha \sim$ 0.10--0.25), the non-dimerization configuration of fully lithiated Li$_2$RuO$_3$ is lower in energy than the dimerization one. In undimerized Li$_2$RuO$_3$, the Ru--Ru bond lengths are calculated to be 2.93 {\AA}, 2.94 {\AA}, and 2.99 {\AA}, and the Ru ions has a local magnetic moment of about 1.6$\mu_{\rm B}$, all obtained in HSE ($\alpha = 0.25$). For comparison, the bond lengths are $\sim$2.5 {\AA} and 3.1 {\AA} and the magnetic moment is $\sim$0.8$\mu_{\rm B}$ in the high-energy, dimerized Li$_2$RuO$_3$.

The fact that the Ru--Ru dimerization configuration of fully lithiated Li$_2$RuO$_3$ is energetically less favorable in our HSE calculations with $\alpha \sim$ 0.10--0.25 is consistent with the absence of Ru--Ru dimers in the vast majority of Li$_2$RuO$_3$ single-crystals.\cite{Wang2014PRB} Our HSE results are also in contrast to those obtained in LDA/GGA, often reported in the literature, in which Ru--Ru dimers are found even in pristine Li$_2$RuO$_3$. The discrepancy can be ascribed to the well-known tendency of LDA/GGA to overdelocalize electrons and hence to favor metal--metal dimerization. 

Note that we do, however, observe Ru--Ru dimerization in Li$_2$RuO$_3$ in the presence of lithium vacancies. In calculations using supercell sizes ranging from unit cells (2 f.u.) to 4$\times$4$\times$1 supercells (32 f.u.), a Ru--Ru dimer is found to form in the vicinity of a void formed by the removal of a lithium. In a 4$\times$4$\times$1 supercell, for example, which corresponds to $\sim$3\% lithium vacancy, the Ru--Ru dimer has a bond length of 2.75 {\AA}, compared to 2.92--3.00 {\AA} of the other Ru--Ru bonds. This calculated value is in excellent agreement with the short Ru--Ru bond of 2.73 {\AA} in some Li$_2$RuO$_3$ single-crystals reported by Wang et al.\cite{Wang2014PRB} Note that Miura et al.\cite{Miura2007JPCJ} reported a slightly smaller value (2.568 {\AA}) for the short Ru--Ru bond in polycrystalline Li$_2$RuO$_3$. All these values are larger than the bond length ($\sim$2.50 {\AA}) of the Ru--Ru dimer in the dimerization (and high-energy) configuration of Li$_2$RuO$_3$.  

Given the fact that the dimerization configuration is higher in energy in fully lithiated Li$_2$RuO$_3$, the experimental observation of the Ru--Ru dimerization in some (nominally) fully lithiated Li$_2$RuO$_3$ samples reported in the literature could be due to the presence of lithium vacancies caused by lithium loss during the preparation of the material at high temperatures. Indeed, Jimenez-Segura {\it et al.}\cite{Jimenez-Segura2016PRB} reported that the dimer formation is sensitive to the preparation procedure and the amount of the RuO$_2$ impurity phase is increased after the repeated grinding and heating steps which indicates lithium loss. It is also possible that both the undimerized and dimerized structures can coexist in a Li$_2$RuO$_3$ sample.

\subsection{Electronic structure vis-\`{a}-vis delithiation mechanism}\label{sec;electronic}

\begin{figure*}%[t]
\vspace{0.2cm}
\includegraphics*[width=0.98\textwidth]{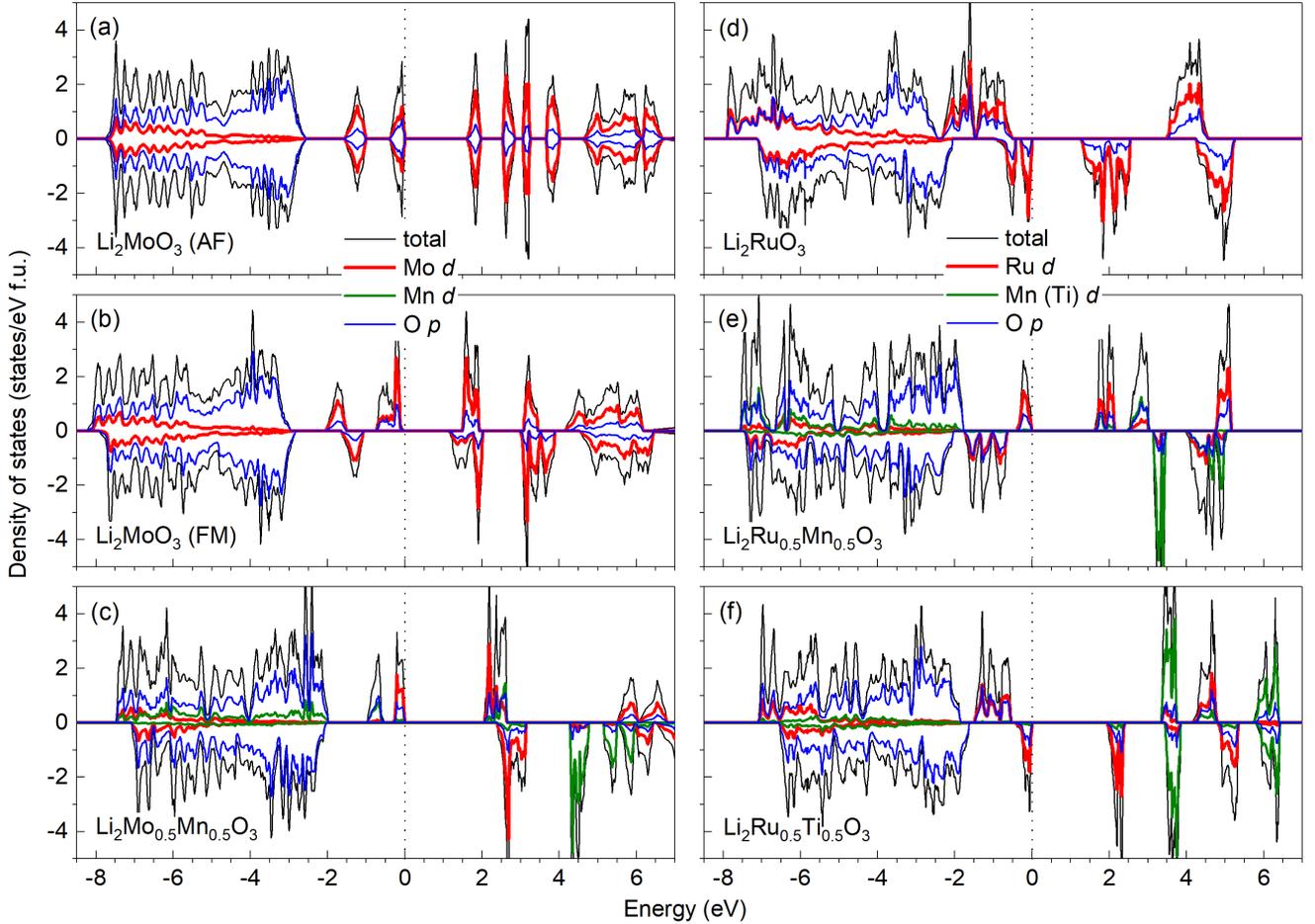}
\vspace{-0.2cm}
\caption{Total and projected electronic density of states of (a) antiferromagnetic and (b) ferromagnetic Li$_2$MoO$_3$, (c) Li$_2$Mo$_{0.5}$Mn$_{0.5}$O$_3$, (d) Li$_2$RuO$_3$, (e) Li$_2$Ru$_{0.5}$Mn$_{0.5}$O$_3$, and (f) Li$_2$Ru$_{0.5}$Ti$_{0.5}$O$_3$ with the majority (minority) spin channel plotted separately on the positive (negative) $y$-axis. The zero of energy is set to the highest occupied states.}
\label{fig;Li2MO3;dos}
\end{figure*}

Figures \ref{fig;Li2MO3;dos}(a) and \ref{fig;Li2MO3;dos}(b) show the electronic density of states (DOS) of Li$_2$MoO$_3$ in the AF and FM spin configurations. The band gap is calculated to be 1.58 eV (AF) or 1.12 eV (FM) within the HSE functional ($\alpha = 0.25$); the gaps are direct in both cases. The electronic structure near the band-gap region are predominantly composed of the Mo $4d$ states. In the AF spin configuration, for example, one Mo atom in Li$_2$MoO$_3$ contributes 62\% to the electronic states at the VBM, whereas each O atom contributes only about 2\%--8\%. Given the feature of the electronic structure of Li$_2$MoO$_3$, oxidation is expected to occur on the transition metal upon delithiation. Indeed, explicit calculations show that, upon removal of a lithium, one of the Mo ions is oxidized to one with a local magnetic moment of $\sim$0$\mu_{\rm B}$, which can be identified as effectively Mo$^{6+}$, consistent with experiments.\cite{Ma2014CM} 

Partially Mn-substituted Li$_2$Mo$_{0.5}$Mn$_{0.5}$O$_3$ is created by replacing one of the two Mo atoms in the unit cell with Mn. In this mixed-metal compound, Mn is found to be stable as high-spin Mn$^{3+}$ ($3d^4$) and Mo as Mo$^{5+}$ ($4d^1$). The change in the charge states of the transition metal ions due to the Mn--Mo interaction is consistent with that previously observed in Mo-doped Li$_2$MnO$_3$.\cite{Hoang2017Li2MnO3} The electronic structure reported in Fig.~\ref{fig;Li2MO3;dos}(c) indicates that the highest valence band is predominantly the Mo $4d$ states whereas the lower valence band is predominantly the Mn $3d$ states. We find that, upon lithium removal, Mo$^{5+}$ is oxidized to Mo$^{6+}$ before Mn$^{3+}$ is oxidized to Mn$^{4+}$, consistent with the arrangement of the Mo$^{5+}$ $4d$ and Mn$^{3+}$ $3d$ bands in the energy spectrum. Note that this arrangement and hence the order in which the redox couples are activated may be dependent on the doping level or, more specifically, the feature of the valence-band top of a specific chemical composition and its atomic arrangement.   

\begin{figure}%[t]%
\vspace{0.2cm}
\includegraphics*[width=0.98\linewidth]{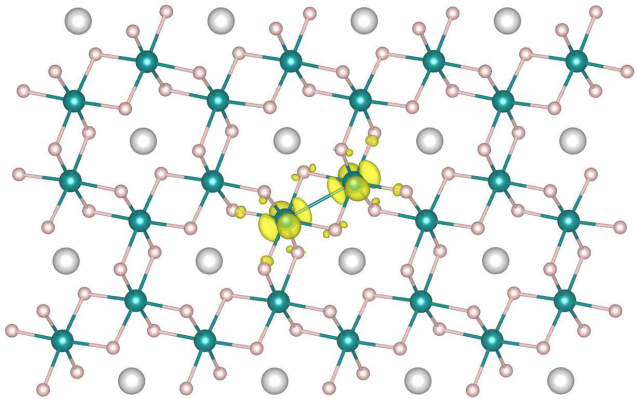}
\caption{The removal of a lithium from the Li$_2$RuO$_3$ supercell results in a negatively charged lithium vacancy (i.e., a void formed by the removal of a Li$^+$ ion, in the Li layer behind the dimer; not shown in the figure), a Ru--Ru dimer, and an electron hole localized on the dimer. The isovalue for the charge-density isosurface (yellow) is set to 0.05 $e$/{\AA}$^3$. Large (gray) spheres are Li, medium (blue) are Ru, and small (red) are O; for clarity, not all the atoms in the supercell are shown.}
\label{fig;dimer}
\end{figure}

\begin{figure*}%[t]%
\vspace{0.2cm}
\includegraphics*[width=0.98\textwidth]{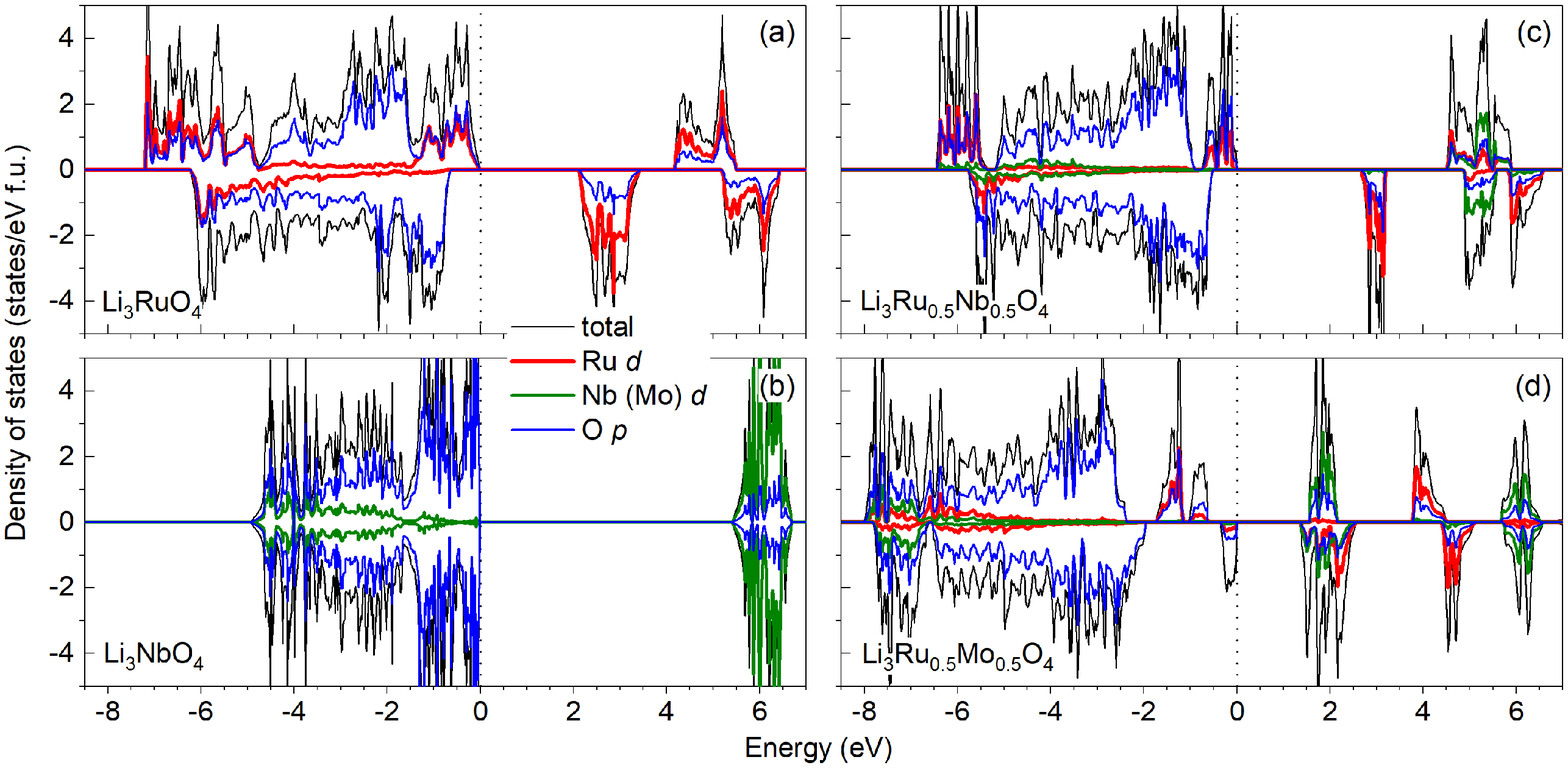}
\vspace{-0.2cm}
\caption{Total and projected electronic density of states of (a) Li$_3$RuO$_4$, (b) Li$_3$NbO$_4$, (c) Li$_3$Ru$_{0.5}$Nb$_{0.5}$O$_4$, and (d) Li$_3$Ru$_{0.5}$Mo$_{0.5}$O$_4$ with the majority (minority) spin channel plotted on the positive (negative) $y$-axis. The zero of energy is set to the highest occupied states.}
\label{fig;Li3MO4;dos}
\end{figure*}

Figures \ref{fig;Li2MO3;dos}(d) show the DOS of Li$_2$RuO$_3$. In this ruthenate, Ru is stable as low-spin Ru$^{4+}$ ($4d^4$). The calculated band gap is 1.27 eV (indirect) within HSE, which appears to be consistent with the ``semiconducting'' behavior reported by Kobayashi {\it et al.}\cite{Kobayashi1995SSI} The electronic structure near the band gap region is predominantly the Ru $t_{2g}^4e_g^0$ states (Note that in an octahedral lattice environment, the five transition-metal $d$-states are split into a lower triplet $t_{2g}$ and an upper doublet $e_g$). Each Ru atom in the cell contributes 40\% to the electronic states at the VBM whereas there is only up to 4\% from each O atom. Upon removal of a lithium, two Ru$^{4+}$ ions in the vicinity of the void left by the removed Li$^+$ ion move closer to each other and form a Ru--Ru dimer with the Ru--Ru bond length of 2.75 {\AA} as discussed in Sec.~\ref{sec;dimerize}. Charge compensation is provided by having an electron hole localized on the Ru--Ru dimer; see Fig.~\ref{fig;dimer}. 

\begin{figure*}%[t]%
\vspace{0.2cm}
\includegraphics*[width=0.98\textwidth]{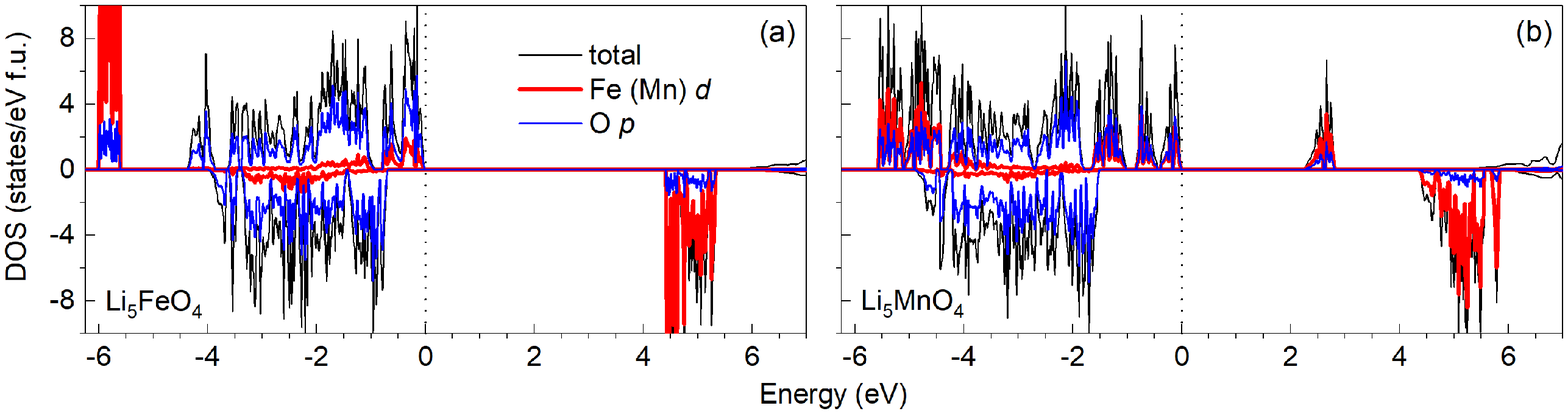}
\vspace{-0.2cm}
\caption{Total and projected electronic density of states (DOS) of (a) Li$_5$FeO$_4$ and (b) Li$_5$MnO$_4$ with the majority (minority) spin channel plotted on the positive (negative) $y$-axis. The zero of energy is set to the highest occupied states.}
\label{fig;Li5MO4;dos}
\end{figure*}

Figures \ref{fig;Li2MO3;dos}(e) and \ref{fig;Li2MO3;dos}(f) show the electronic structure of Li$_2$Ru$_{0.5}$Mn$_{0.5}$O$_3$ and Li$_2$Ru$_{0.5}$Ti$_{0.5}$O$_3$. In the partially Mn-substituted compound, Mn is stable as Mn$^{4+}$ ($3d^3$). The Mn $3d$ states are high up in the conduction band, whereas the top of the valence band is predominantly the Ru $4d$ states. The Mn ion is thus electrochemically inactive in Li$_2$Ru$_{0.5}$Mn$_{0.5}$O$_3$, similar to that in Li$_2$MnO$_3$.\cite{Hoang2015PRA} In the partially Ti-substituted compound, Ti is stable as Ti$^{4+}$ ($3d^0$) and the Ti $3d$ states are also high up in the conduction band. The electronic structure of Li$_2$Ru$_{0.5}$Sn$_{0.5}$O$_3$ (not shown in the figure) is very similar to that of Li$_2$Ru$_{0.5}$Sn$_{0.5}$O$_3$. As far as the delithiation mechanism is concerned, these partially Mn-, Ti-, and Sn-substituted compounds are similar to the parent compound Li$_2$RuO$_3$ in which Ru is the electrochemically active center, at least in the early stages of delithiation. 

Figure \ref{fig;Li3MO4;dos}(a) shows the electronic structure of Li$_3$RuO$_4$. In this compound, Ru is stable as Ru$^{5+}$ ($4d^3$) with a calculated magnetic moment of 2.08 $\mu_{\rm B}$. The calculated band gap is 2.11 eV, an indirect gap. The electronic structure near the band-gap region is predominantly composed of the Ru $t_{2g}^3e_g^0$ states. Each Ru atom in the primitive contributes about 23\% to the states at the VBM, where each O atom only contributes about 5--8\%. We find that, upon delithiation, oxidation occurs mainly on the Ru ion with Ru$^{5+}$ oxidized to what can be identified as Ru$^{6+}$ ($4d^2$) with a calculated magnetic moment of $\sim$1.4$\mu_{\rm B}$, which is consistent with the fact that the highest occupied states in the electronic structure of Li$_3$RuO$_4$ are predominantly composed of the Ru $4d$ states. 

The electronic structure of Li$_3$NbO$_4$, on the other hand, is characterized by having predominantly the O $2p$ states at the valence-band top and the Nb $4d$ states at the conduction-band bottom; see Fig.~\ref{fig;Li3MO4;dos}(b). The calculated band gap is 5.39 eV (direct). Upon delithiation, oxidation occurs on oxygen, turning O$^{2-}$ in Li$_3$NbO$_4$ into O$^-$ (i.e., a localized hole on oxygen). The delithiation mechanism in this Li-rich oxide thus involves anionic redox.

Figures \ref{fig;Li3MO4;dos}(c) and \ref{fig;Li3MO4;dos}(d) shows the electronic structure of the partially Nb- and Mo-substituted compounds. In Li$_3$Ru$_{0.5}$Nb$_{0.5}$O$_4$, Nb is stable as Nb$^{5+}$ ($4d^0$). The electronic structure near the band gap region is predominantly Ru $4d$ states; the Nb $4d$ states are high up in the conduction band. Upon delithiation, oxidation will therefore occur on Ru whereas Nb is electrochemically inactive. In Li$_3$Ru$_{0.5}$Mo$_{0.5}$O$_4$, Ru and Mo are stable as Ru$^{4+}$ ($4d^4$) and Mo$^{6+}$ ($4d^0$), respectively. There is thus charge transfer between the two transition-metal ions. The Mo $4d$ states are in the conduction band, whereas the top of the valence band is predominantly composed of the Ru $4d$ states. Upon lithium removal, Ru$^{4+}$ is oxidized to Ru$^{5+}$ during the early stages of delithiation.

Figures \ref{fig;Li5MO4;dos}(a) and \ref{fig;Li5MO4;dos}(b) show the electronic structure of Li$_5$FeO$_4$ and Li$_5$MnO$_4$. In these compounds, the transition metal is tetrahedrally coordinated with oxygen with the five transition-metal $d$-states split into a lower doublet $e$ and an upper triplet $t_2$. Iron in Li$_5$FeO$_4$ is stable as high-spin Fe$^{3+}$ ($3d^5$) with a calculated magnetic moment of 4.04$\mu_{\rm B}$. The calculated band gap is 4.41 eV (direct). The electronic structure near the band gap region is composed of the Fe $e^2 t_2^3$ and O $2p$ states. A detailed analysis shows that each Fe atom contributes 3.3\% to the electronic states at the VBM, whereas each O atom contributes about 2.1--2.5\%. In Li$_5$MnO$_4$, Mn is stable as Mn$^{3+}$ ($3d^4$) with a calculated magnetic moment of 3.62$\mu_{\rm B}$. The electronic structure near the band gap region is composed of the Mn $e^2 t_2^2$ and O $2p$ states and the calculated band gap is 2.25 eV (direct). Each Mn atom contributes 5.3\% to the electronic states at the VBM, whereas each O atom contributes 1.1--2.1\%. In both compounds, the first stage of delithiation (i.e., the removal of the first lithium from the unit cell) is associated with the oxidation of Fe$^{3+}$ (Mn$^{3+}$) to Fe$^{4+}$ (Mn$^{4+}$). Later stages are expected to involve oxidation of both the transition metal and oxygen. It has been reported that both cationic and anionic redox occur in Li$_5$FeO$_4$.\cite{Zhan2017NE}  

\section{Conclusions} 

We have carried out a hybrid density-functional study of the atomic and electronic structure of select Li-rich complex oxide battery electrode materials. The calculated lattice parameters are in good agreement with experiments. Dimerization of the Mo ions is observed in layered Li$_2$MoO$_3$, even in the fully lithiated compound. In layered Li$_2$RuO$_3$, Ru--Ru dimerization occurs only upon lithium removal or when the material is Li-deficient, in contrast to what is commonly believed that the dimerization occurs in the fully lithiated compound. There is a reduction in the local magnetic moments associated with the dimerization. In light of an analysis of the calculated electronic structure, we have discussed the delithiation mechanism in Li$_2$MoO$_3$, Li$_2$RuO$_3$, Li$_3$RuO$_4$, Li$_3$NbO$_4$, Li$_5$FeO$_4$, Li$_5$MnO$_4$ and their derivatives.

\begin{acknowledgments}

This work was supported financially by the National Institute of General Medical Sciences of the National Institute of Health under Award No.~R15GM122063 and made use of computing resources at the Center for Computationally Assisted Science and Technology (CCAST) at North Dakota State University. 

\end{acknowledgments}

%\bibliography{batterymaterials}
%
\end{document}